\documentclass[aps,prd,reprint,nofootinbib,superscriptaddress,floatfix]{revtex4-2}

\usepackage[T1]{fontenc}
\usepackage[utf8]{inputenc}
\usepackage{lmodern}
\usepackage{amsmath,amssymb,amsfonts,bm}
\usepackage{graphicx}
\graphicspath{{../figures/}{./figures/}}
\usepackage{microtype}
\usepackage{booktabs}
\usepackage{array}
\usepackage[hidelinks]{hyperref}

\hypersetup{
  pdftitle={From cross-section degeneracies to phase-sensitive observables in near-threshold J/psi production},
  pdfauthor={Arkadiy I. Syamtomov},
  pdfsubject={Near-threshold J/psi production, scalar and spin-two amplitudes, non-forward OPE, GPDs and polarization observables},
  pdfkeywords={J/psi production, non-forward OPE, scalar production direction, spin-two amplitude, final-state interactions, electroproduction}
}

\newcommand{\GeV}{\mathrm{GeV}}
\newcommand{\paramvec}{(M_{A_g},C_g(0),M_{C_g},M_{A_q},C_q(0),M_{C_q})}

\begin{document}

\title{From cross-section degeneracies to phase-sensitive observables in near-threshold \(J/\psi\) production}
\author{Arkadiy I.\ Syamtomov}
\email{arkady.syamtomov@gmail.com}
\affiliation{Bogolyubov Institute for Theoretical Physics, National Academy of Sciences of Ukraine, Kyiv, Ukraine}

\begin{abstract}
Near-threshold \(J/\psi\) production is sensitive to scalar and traceless spin-two chromoelectric structure in the nucleon.  We study whether present integrated and differential measurements can separate these contributions and constrain their relative phase.  The GlueX integrated cross section does not determine a unique scalar contribution: production and elastic scalar terms are strongly correlated, while flatter non-forward profiles drive the fitted scalar strength to zero.  In a 38-point large-\(\xi\) differential analysis, the scalar term is collinear at fixed kinematics with the retained \(C\)-GFF direction, leading to nearly degenerate solutions and strong profile dependence.  A common single-channel final-state factor preserves the scalar--spin-two relative phase, whereas coupled-channel rescattering can modify it.  Its determination therefore requires polarization-sensitive observables together with process-specific complex helicity amplitudes.
\end{abstract}

\maketitle

\section{Introduction}

The small size of the $J/\psi$ makes heavy-quarkonium production a
particularly direct probe of gluonic fields in the nucleon.  In the
short-distance treatment of ~\cite{Peskin1979,BhanotPeskin1979}, the interaction
of a compact heavy $Q\bar Q$ state with soft hadronic matter is
organized by chromoelectric multipoles.
At leading $E1\times E1$ order the relevant chromoelectric response can,
in a fixed renormalized operator basis, be separated into scalar and
traceless spin-two components of the QCD energy-momentum tensor.  The
former contains the trace-anomaly and sigma-term contributions, whereas
the latter is related to twist-two gluon structure.  This connection has motivated studies of quarkonium--nucleon interactions
and near-threshold production as probes of gluonic nucleon structure
~\cite{KharzeevSatzSyamtomovZinovjev1999,Hoodbhoy1999,HattaYang2018,GuoJiLiu2021}.  Near-threshold OPE constraints and their connection to
the higher-energy diffractive regime are discussed in
~\cite{Syamtomov2026OPEHighEnergy}.

The interpretation becomes nontrivial because physical threshold
kinematics are strongly non-forward.  The forward OPE/PDF input fixes
the twist-two amplitude at $t=0$, whereas exact $J/\psi$
photoproduction threshold occurs at $t\simeq-2.2~\GeV^2$.
Moreover, the physical $t$ interval changes rapidly with $W$ near
threshold.  The assumed non-forward profile therefore affects not only
the normalization but also the energy dependence of the
$t$-integrated cross section.  For an exponential profile
$F(t)\sim\exp(bt/2)$, a larger $b$ gives a steeper fall with $|t|$,
whereas a smaller $b$ gives a flatter continuation.  As shown below, a scalar term added at Born level to the production kernel and the corresponding $K$-matrix fit with $C_S^{(K)}=0$ both change the fixed steep-profile fit only moderately.  When the production and elastic scalar coefficients are varied independently, the same nine points allow strongly correlated pairs of coefficients rather than a unique scalar extraction.  The scalar contribution also vanishes when the non-forward profile is allowed to become flatter.  The integrated cross section
therefore does not by itself distinguish a scalar contribution from the assumed
non-forward continuation or from the allocation of scalar strength between
production and final-state rescattering.

Recent GlueX, $J/\psi$-007 and CLAS12 measurements provide differential
information in the near-threshold region
~\cite{GlueX2023,Duran2023,CLAS122026}, making it possible to ask
whether retaining the momentum-transfer dependence resolves this
ambiguity.  This question is especially relevant to interpretations of
near-threshold production in terms of gluon GPDs and gravitational form
factors~\cite{GuoJiLiu2021,GuoJiLiuYang2023,GuoJiYuan2024,PentchevChudakov2025}.  We find that differential information alone
does not provide a model-independent separation.  In the retained
large-$\xi$ GPD representation, the phenomenological scalar production
term and the $C$-GFF contribution occupy the same scalar Dirac direction
at fixed kinematics and differ only through their assumed $(t,\xi)$
profiles.  The resulting scalar coordinate consequently depends on the chosen
profiles and allowed parameter ranges rather than defining an independently
identified nucleon form factor.

A second limitation concerns phase information.  For a fixed physical
helicity or partial-wave channel, write the chromoelectric part of the
production amplitude as
\[
 {\cal F}_{\rm CE}={\cal F}_0+{\cal F}_2 .
\]
The unpolarized rate contains
\[
 |{\cal F}_{\rm CE}|^2
 =|{\cal F}_0|^2+|{\cal F}_2|^2
 +2\,\mathrm{Re}({\cal F}_0^*{\cal F}_2),
\]
and therefore depends on modulus-squared and real-interference combinations, but
does not supply the signed imaginary interference required to fix
the relative phase.  We examine how this phase is affected by
single- and coupled-channel final-state interactions and under what
conditions polarization observables can provide the missing
interference quadrature.  The connection of such an observable to a
scalar--spin-two phase is necessarily conditional on the
process-specific complex production matching.

The integrated, differential and phase-sensitive analyses are therefore treated here as a single unified identifiability study.  We proceed from the ambiguity of the integrated cross section, through the additional but still incomplete information retained in present differential data, to the phase information that can in principle be supplied by polarization-sensitive measurements.

\section{Scalar and spin-two production near threshold}
\label{sec:production}

It is useful to distinguish the chromoelectric nucleon source from its
production and final-state realization,
\begin{equation}
 N_a\longrightarrow {\cal M}^{J}_{a,\lambda'\lambda}
 \longrightarrow {\cal F}^{J}_{a,\lambda'\lambda},
 \qquad a=0,2,
\label{eq:hierarchy}
\end{equation}
where $N_a$ is the renormalized chromoelectric source,
${\cal M}^{J}_{a,\lambda'\lambda}$ the corresponding matched
$\gamma^{(*)}N\to J/\psi N$ production amplitude before strong
rescattering, and ${\cal F}^{J}_{a,\lambda'\lambda}$ its
final-state-dressed contribution.  We use the renormalized
scalar/spin-two basis and matching convention of
~\cite{Syamtomov2026EMT,Syamtomov2026Matching}.
Finite-transverse-momentum scalar and spin-two response profiles based
on this operator separation have been analyzed in
~\cite{Syamtomov2026LFProfiles}; here we ask how the non-forward
structure affects their separation in production amplitudes and their
relative phase.

We denote by $N_0$ and $N_2$ the reduced scalar and traceless
spin-two chromoelectric nucleon responses.  In the forward normalization
of ~\cite{Syamtomov2026EMT},
\begin{align}
 N_0(0)&=C_{\cal A}\,[M-\Sigma_N]+C_\sigma\,\Sigma_N,\notag\\
 N_2(0)&=\frac{3M}{4}
 \left[C_q^{(2)}A_q+C_g^{(2)}A_g\right].
\label{eq:N0N2}
\end{align}
Here $\Sigma_N$ denotes the nucleon sigma-term contribution, while
$C_{\cal A}$, $C_\sigma$, $C_q^{(2)}$ and $C_g^{(2)}$ are the
renormalized source-level scalar and traceless spin-two matching
coefficients in the convention of ~\cite{Syamtomov2026EMT}.
The underlying mass/trace-anomaly organization follows the standard
QCD energy-momentum-tensor decomposition
~\cite{Ji1995Mass,HattaRajanTanaka2018}.

At leading $E1\times E1$ order and leading order in $1/m_c$, the $1S$
quarkonium response is a scalar in heavy-spin space and isotropic in
the dipole indices~\cite{LukeManoharSavage1992,
BrambillaPinedaSotoVairo2000}.  Before embedding this response into the
electromagnetic production amplitude, it is useful to isolate the
heavy-spin structure of the forward quarkonium--nucleon matching.
Denoting this auxiliary matched response by ${\cal R}_a$, its leading
forward form is
\[
 {\cal R}_a^{\,is,js'}(0)
 =r_aN_a(0)\,\delta^{ij}\delta_{ss'},
 \qquad a=0,2,
\]
where $a=0,2$ labels the scalar and traceless spin-two operator sectors,
$i,j$ are the Cartesian spin-1 polarization indices of the $J/\psi$,
and $s,s'$ are nucleon spin indices.  The coefficient $r_a$ denotes the
leading quarkonium matching factor.  Corrections beyond this leading
forward limit include heavy-spin-breaking, higher-multipole,
derivative and finite-$t$ tensor structures.

Rotational invariance alone would, by Schur's lemma, allow independent
coefficients in the $J=\tfrac12$ and $J=\tfrac32$ irreducible
subspaces of $1\otimes\tfrac12=\tfrac12\oplus\tfrac32$.  The leading
$E1\times E1$ response is more restrictive: the forward tensor above
is proportional to the identity on the full product spin space.
Projection onto the two $S$-wave total-spin channels therefore gives
\begin{equation}
 {\cal R}_a^{{}^{2}S_{1/2}}(0)
 ={\cal R}_a^{{}^{4}S_{3/2}}(0)
 =r_aN_a(0),
 \qquad a=0,2.
\label{eq:same_projection}
\end{equation}
Thus the equality of the two forward $J/\psi N$ spin channels follows
from the leading spin-independent $E1\times E1$ structure, while the
scalar and spin-two operator sectors remain distinct.

Physical photoproduction is already strongly non-forward at threshold.
Kinematically,
\begin{equation}
 t_{\rm thr}=-\frac{M M_\psi^2}{M+M_\psi}
 \simeq -2.23~\GeV^2,
\label{eq:t_threshold}
\end{equation}
while $t_{\min}\simeq-1.30~\GeV^2$ at $W=4.10$~GeV and
$-0.93~\GeV^2$ at $W=4.20$~GeV.  A derivative expansion with
non-forward scale $\Lambda_t$ is governed parametrically by
$|t|/\Lambda_t^2$.  At exact threshold,
\[
 \frac{|t_{\rm thr}|}{\Lambda_t^2}
 \simeq
 0.56\left(\frac{2~\GeV}{\Lambda_t}\right)^2.
\]
For orientation, requiring this expansion parameter to be below
$0.3$ gives $\Lambda_t>2.73~\GeV$, while the weaker condition below
$0.5$ gives $\Lambda_t>2.11~\GeV$.  The finite-$t$ expansion is
therefore controlled only if the relevant non-forward scale is
sufficiently hard; otherwise corrections at physical threshold can be
large.

For each component $a=0,2$ and a fixed helicity or partial-wave
projection, the production step may be written as
\begin{equation}
 {\cal M}_{a,\lambda}(W,t)
 =\kappa_{a,\lambda}(W,t;\mu)N_a(t;\mu)
 +{\cal M}^{\rm sub}_{a,\lambda}(W,t;\mu).
\label{eq:operator_matching}
\end{equation}
Here $\lambda$ is a collective label for the chosen fixed helicity or
partial-wave projection.  The process-dependent coefficient
$\kappa_{a,\lambda}(W,t;\mu)$ matches the renormalized nucleon source
$N_a(t;\mu)$ onto the corresponding electromagnetic production
amplitude before strong final-state rescattering; it contains the
short-distance quarkonium-production dynamics, kinematic and helicity
projection, and the compensating renormalization-scale dependence.
The remainder ${\cal M}^{\rm sub}_{a,\lambda}$ collects production
structures not represented by the chosen local matrix element.
Consequently, identifying a phenomenological scalar production term
directly with the source $N_0$ requires additional production-matching
assumptions.

At an isolated timelike $J/\psi$ pole, the electromagnetic-current
residue factorizes into the current coupling and the full on-shell
$J/\psi N$ scattering amplitude
~\cite{BauerSpitalYenniePipkin1978}.  For a fixed polarization,
\[
 {\cal F}^{\gamma^*,\mathrm{pole}}_{a,\lambda}
 =
 R_{\gamma\psi,\lambda}(q^2)\,
 {\cal T}^{\psi N}_{a,\lambda},
 \qquad a=0,2,
\]
where ${\cal T}^{\psi N}_{a,\lambda}$ denotes the component of the
on-shell $J/\psi N\to J/\psi N$ amplitude associated with the same
fixed scalar/spin-two decomposition.  The electromagnetic pole factor
$R_{\gamma\psi,\lambda}$ is independent of the operator label $a$ and
therefore cancels polarization by polarization,
\begin{equation}
 \frac{{\cal F}^{\gamma^*,\mathrm{pole}}_{2,\lambda}}
      {{\cal F}^{\gamma^*,\mathrm{pole}}_{0,\lambda}}
 =
 \frac{{\cal T}^{\psi N}_{2,\lambda}}
      {{\cal T}^{\psi N}_{0,\lambda}}.
\label{eq:pole_ratio}
\end{equation}
Pole factorization preserves a chosen linear decomposition of the
residue but does not define it; reduction of
Eq.~\eqref{eq:pole_ratio} to $N_2/N_0$ requires additional matching
assumptions.  Away from the pole the full electromagnetic amplitude
contains, in general, a part regular in $q^2$,
\[
 {\cal F}^{\gamma^*}_{a,\lambda}(q^2)
 =R_{\gamma\psi,\lambda}(q^2){\cal T}^{\psi N}_{a,\lambda}
 +{\cal F}^{\rm reg}_{a,\lambda}(q^2).
\]
Identifying the real-photon amplitude at $q^2=0$ with the pole term
alone is the vector-dominance assumption.  Direct production, higher
vector states and continuum contributions are regular at the
$J/\psi$ pole but need not be negligible at the photon point, and need
not preserve the scalar/spin-two ratio of the pole residue
~\cite{HufnerKopeliovich1998,
DuBaruGuoHanhartMeissnerNefedievStrakovsky2020}.

Direct production also carries nontrivial Lorentz, helicity and
partonic-momentum dependence, so the scalar and traceless spin-two
coefficient structures need not be proportional
~\cite{SunTongYuan2022,GuoJiYuan2024}.  Denoting the electromagnetic
production current by
\[
 {\cal H}^{\mu}_{\lambda_\psi;s's}
 \equiv
 \langle J/\psi(p_\psi,\lambda_\psi),N(p',s')|
 J_{\rm em}^{\mu}(0)|N(p,s)\rangle ,
\]
its operator matching has schematically the form
\[
 {\cal H}^{\mu}_{\lambda_\psi;s's}
 =\epsilon_{\psi}^{*\nu}(\lambda_\psi)
 \left[
 C_{0\,\nu}^{\ \mu}\langle{\cal O}_0\rangle
 +C_{2\,\nu}^{\ \mu,\alpha\beta}
   \langle O^{(2)}_{\alpha\beta}\rangle
 \right]+\cdots .
\]
Here $\langle{\cal O}_0\rangle$ and
$\langle O^{(2)}_{\alpha\beta}\rangle$ denote the scalar and traceless
spin-two nucleon matrix elements whose reduced forward combinations
define $N_0$ and $N_2$, respectively.  Electromagnetic current
conservation gives
\[
 q_\mu{\cal H}^{\mu}=0.
\]
The scalar and spin-two terms can therefore acquire different
kinematic and helicity projections even when their underlying forward
nucleon matrix elements are real.

After helicity and partial-wave projection we define the production
ratio and phase by
\[
 \eta_{{\rm prod},\lambda'\lambda}^{J}
 =\frac{{\cal M}^{J}_{2,\lambda'\lambda}}
        {{\cal M}^{J}_{0,\lambda'\lambda}},
 \qquad
 \phi_{20,{\rm prod},\lambda'\lambda}^{J}
 =\arg\eta_{{\rm prod},\lambda'\lambda}^{J}.
\]
For the final-state-dressed amplitudes we define the distinct ratio
and phase
\[
 \eta_{{\rm phys},\lambda'\lambda}^{J}
 =\frac{{\cal F}^{J}_{2,\lambda'\lambda}}
        {{\cal F}^{J}_{0,\lambda'\lambda}},
 \qquad
 \phi_{20,{\rm phys},\lambda'\lambda}^{J}
 =\arg\eta_{{\rm phys},\lambda'\lambda}^{J}.
\]
In general $\eta_{\rm phys}^{J}\neq\eta_{\rm prod}^{J}$ because
final-state rescattering can dress the two production components
differently; equality follows only when they acquire the same
final-state factor.  Different helicity channels can therefore carry
different production and physical ratios and phases.  Cross-section
observables constrain bilinears of the final-state amplitudes
${\cal F}$; translating them into production-level ratios, or
ultimately into $N_2/N_0$, requires assumptions about the matching and
final-state dynamics.  We first examine how much of this separation is
identifiable from present integrated and differential data before
turning to the phase information.

\section{Constraints from photoproduction data}
\label{sec:data}

\subsection{Integrated cross-section constraints}

We first consider the integrated near-threshold cross section.  Since the momentum-transfer dependence is integrated out, a scalar production term can be correlated with the assumed non-forward behavior of the elastic amplitude.  Near threshold the production partial waves obey ${\cal M}_{\ell}=O(k_\psi^\ell)$, so the $\ell=0$ channel is dominant.  Here $S$ wave denotes orbital angular momentum $\ell=0$, while the scalar operator sector is labelled separately by $a=0$.  For the scalar production profile we use
\[
 G_S(t)=\frac{M}{(1-t/m_s^2)^2},\qquad m_s=1.24~\GeV,
\]
corresponding to the central GlueX fit of~\cite{Kharzeev2021MassRadius}, and define $\bar G_S(s)=\tfrac12\int_{-1}^{1}dz\,G_S(t(s,z))$.  The target-mass-corrected OPE input of~\cite{Syamtomov2026OPEHighEnergy} supplies the reference production amplitude.

At Born level the scalar term enters as
\begin{equation}
 P_{\ell=0}(s)=P_{\ell=0}^{\rm OPE}(s)+C_S^{(P)}\bar G_S(s).
\label{eq:integrated_production_only}
\end{equation}
To allow distinct scalar contributions in production and elastic rescattering, we use the single-channel $K$-matrix form~\cite{Aitchison1972}
\begin{equation}
 {\cal A}_{\ell=0}^{(K)}(s)=
 \frac{P_{\ell=0}^{\rm OPE}(s)+C_S^{(P)}\bar G_S(s)}
 {1-i\rho_{\psi N}(s)\,[K_{\ell=0}^{\rm emb}(s)+C_S^{(K)}\bar G_S(s)]},
\label{eq:integrated_kmatrix}
\end{equation}
with $\rho_{\psi N}=2k_\psi/\sqrt{s}$.  The coefficients $C_S^{(P)}$ and $C_S^{(K)}$ parameterize scalar contributions to the reduced production and elastic amplitudes, respectively; we write them below as $C_P$ and $C_K$.  The real kernel $K^{\rm emb}$ has its forward normalization fixed by the OPE input, while its non-forward continuation is a phenomenological input.

Writing
\[
 N=P_{\ell=0}^{\rm OPE}+C_P\bar G_S,\qquad
 D=K_{\ell=0}^{\rm emb}+C_K\bar G_S,
\]
the threshold behavior is
\begin{align}
 |{\cal A}^{(K)}|^2&=N^2[1-\rho_{\psi N}^2D^2+O(\rho_{\psi N}^4)],\notag\\
 \arg {\cal A}^{(K)}&=\arg N+\rho_{\psi N}D+O(\rho_{\psi N}^3).
\label{eq:integrated_threshold_expansion}
\end{align}
Thus the elastic contribution first changes the unpolarized modulus at $O(k_\psi^2)$, while the production scalar remains in the finite threshold numerator.  The physical photoproduction cross section still vanishes at threshold through phase space.

To specify the non-forward elastic kernel we write
\begin{align}
 K^{\rm emb}(s,t)&=K^{\rm emb}(s,0)F_K(t),\notag\\
 \overline F_K(s)&=\frac12\int_{-1}^{1}dz\,F_K(t(s,z)),
\end{align}
so that $K_{\ell=0}^{\rm emb}(s)=K^{\rm emb}(s,0)\overline F_K(s)$ with $F_K(0)=1$.  We compare the diffractive exponential $F_{\rm exp}=e^{bt/2}$ with dipole continuations $F_{\rm dip}=(\Lambda^2/(\Lambda^2-t))^2$ motivated by two-gluon form-factor phenomenology~\cite{FrankfurtStrikman2002}.

For the nine GlueX-2023 points with $W\le4.55~\GeV$~\cite{GlueX2023},
\begin{equation}
 \chi^2_{\rm int}=\sum_{i=1}^{9}
 \frac{[\sigma_i^{\rm data}-\sigma_i^{\rm model}]^2}{(\delta\sigma_i)^2}.
\label{eq:integrated_calibrated_stat}
\end{equation}
The fit uses the continuous positive-$D$ convention; the equivalent opposite-sign representation is given in Appendix~\ref{app:integrated_representation}.  The principal results are summarized in Table~\ref{tab:integrated_scalar_placement}.

\begin{table*}[tbp]
\centering
\scriptsize
\setlength{\tabcolsep}{3.4pt}
\caption{Scalar placement in the integrated nine-point fit using the positive-$D$ convention.  The Born and $K$-matrix results are compared with the corresponding calculations in which the scalar terms are omitted.}
\label{tab:integrated_scalar_placement}
\begin{ruledtabular}
\begin{tabular}{lrrrrr}
Treatment & $C_S^{(P)}$ & $C_S^{(K)}$ & $\chi^2_{\rm int}$ & $\Delta\chi^2_{\rm Born}$ & $\Delta\chi^2_{K}$\\
\hline
OPE/TMC Born & 0 & -- & 49.185 & 0 & --\\
Born production scalar, no FSI & 6.848 & -- & 45.428 & 3.757 & --\\
$K$-matrix, $C_S^{(P)}=C_S^{(K)}=0$ & 0 & 0 & 51.561 & -- & 0\\
production scalar, $C_S^{(K)}=0$ & 8.224 & 0 & 46.278 & -- & 5.283\\
elastic scalar only, $C_S^{(P)}=0$ & 0 & 7.639 & 45.671 & -- & 5.889\\
shared $C_S^{(P)}=C_S^{(K)}$, fitted & 12.689 & 12.689 & 28.228 & -- & 23.332\\
independent $C_S^{(P)},C_S^{(K)}$ & 61.142 & 22.263 & 0.500 & -- & 51.060\\
\end{tabular}
\end{ruledtabular}
\end{table*}

Adding a scalar term only to the production amplitude gives a modest improvement, both at Born level and after inclusion of the elastic denominator.  When $C_P$ and $C_K$ are varied independently, however, they become strongly correlated and widely different pairs describe the nine points with similar quality; allowing the overall normalization to vary strengthens this correlation.  The detailed coefficient-range dependence and the coherent-sign mapping are given in Appendix~\ref{app:integrated_representation}.

The second ambiguity concerns the non-forward continuation itself.  Harder profiles reproduce the observed energy dependence substantially better and, in the shared-coefficient model, drive the fitted scalar coefficient to zero.  The scalar length scale is correspondingly weakly constrained.  Figure~\ref{fig:integrated_profile} summarizes this dependence, while the complete numerical comparison is given in Appendix~\ref{app:integrated_representation}.

Together these results show that the nine integrated points determine neither a unique scalar production strength, its allocation to elastic rescattering, nor the non-forward kernel shape.

\begin{figure}[tbp]
\centering
\includegraphics[width=\columnwidth]{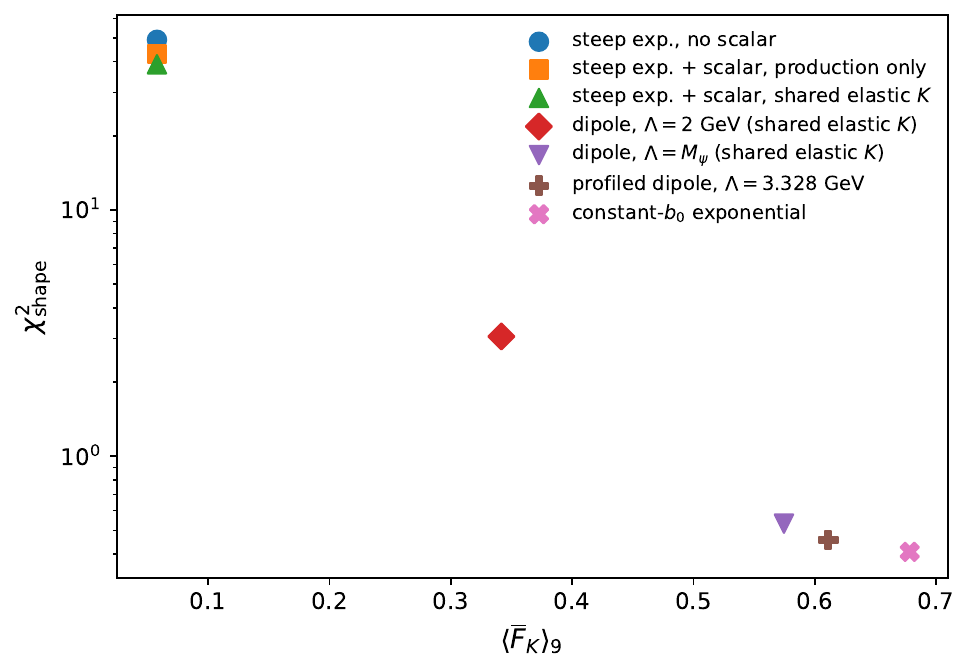}
\caption{Dependence of the integrated nine-point fit on the assumed non-forward elastic-kernel profile.  The horizontal coordinate is the nine-point mean of the projected kernel profile, $\langle\overline F_K\rangle_9$.  The three steep-exponential points correspond to the curves with the scalar term omitted, with a Born scalar term, and with a common scalar coefficient in the production and elastic terms at $C_0=6.45924$, with one overall normalization factor fitted for each curve.  Harder dipole and constant-slope continuations drive the shared scalar coefficient to zero.}
\label{fig:integrated_profile}
\end{figure}

The integrated and differential analyses use different representations of the non-forward amplitude.  In the integrated case the elastic kernel is continued away from the forward OPE input, whereas the differential analysis uses the large-$\xi$ GPD representation and retains the measured $t$ dependence explicitly.  The scalar profile used below is therefore independent of the $m_s=1.24~\GeV$ form adopted for the integrated cross section.

\subsection{Differential scalar--GFF identifiability}

For the differential analysis we use the real leading-conformal-moment truncation of the NLO large-$\xi$ framework of Guo, Yuan and Zhao~\cite{GuoYuanZhao2025}, with
\[
 x=\frac{M_\psi^2-t}{W^2-M^2},\qquad \xi=\frac{x}{2-x},
\]
and the cut $\xi>0.5$.  This retains 12 GlueX, 21 $J/\psi$-007 and 5 CLAS12 measurements, 38 points in total~\cite{Duran2023,CLAS122026,GlueXHEPData2023}.  Relativistic corrections can be important near threshold~\cite{BlaskFlemingMehenRoyStewartZhao2026}; the analysis therefore addresses identifiability within the retained large-$\xi$ representation.

Following~\cite{GuoYuanZhao2025}, we use
\begin{align}
 A_i(t)&=A_i(0)\left(1-\frac{t}{M_{A_i}^2}\right)^{-2},\notag\\
 C_i(t)&=C_i(0)\left(1-\frac{t}{M_{C_i}^2}\right)^{-3},
\end{align}
with $A_g(0)=0.40$, $A_q(0)=0.60$ and $B_g=B_q=0$.  At $\mu_F=m_c$ and $\alpha_s=0.30$, the retained real CFFs may be written
\begin{align}
 {\cal H}(t,\xi)&={\cal H}_A(t,\xi)+{\cal C}_{\rm eff}(t),\notag\\
 {\cal E}(t,\xi)&=-{\cal C}_{\rm eff}(t),\notag\\
 {\cal H}_A(t,\xi)&=\frac{1}{\xi^2}
 \left[\frac85\bar C_g A_g(t)+2\bar C_q A_q(t)\right],\notag\\
 {\cal C}_{\rm eff}(t)&=\frac{32}{5}\bar C_g C_g(t)+8\bar C_q C_q(t),
\label{eq:effective_c_direction}
\end{align}
where $\bar C_g=5/4-0.369\alpha_s$ and $\bar C_q=-0.891\alpha_s$.

The scalar ansatz enters through
\begin{equation}
 {\cal H}\to{\cal H}+{\cal S},\qquad
 {\cal E}\to{\cal E}-{\cal S}.
\label{eq:scalar_cff_shift}
\end{equation}
Equations~\eqref{eq:effective_c_direction} and~\eqref{eq:scalar_cff_shift} show that, at fixed kinematics, the scalar term and the retained $C$-GFF combination occupy exactly the same $(\delta{\cal H},\delta{\cal E})\propto(1,-1)$ Dirac direction.  This collinearity is the central identifiability result of the differential analysis: at one kinematic point the retained observables depend only on the sum ${\cal C}_{\rm eff}+{\cal S}$.  Separation over the measured sample can arise only from the different $(t,\xi)$ dependences assigned to the two terms.

The fitted differential cross section is
\begin{align}
 \frac{d\sigma}{dt}&={\cal N}(W)|G(t,\xi)|^2,\notag\\
 |G|^2={}&(1-\xi^2)|{\cal H}+{\cal E}|^2
 -2\,\mathrm{Re}[{\cal E}^*({\cal H}+{\cal E})]\notag\\
 &+\left(1-\frac{t}{4M^2}\right)|{\cal E}|^2.
\label{eq:gpd_bilinear}
\end{align}
with the normalization ${\cal N}(W)$ of Ref.~\cite{GuoYuanZhao2025}.  We parameterize the scalar term around $t_*=-3.29~\GeV^2$, $\xi_*=0.5684$ as
\begin{align}
 {\cal S}(t,\xi)&=S_* f_S(t,\xi),\notag\\
 f_S(t,\xi)&=
 \left[\frac{1-t/M_S^2}{1-t_*/M_S^2}\right]^{-p_S}
 \left(\frac{\xi}{\xi_*}\right)^{\nu_S},
\qquad f_S(t_*,\xi_*)=1.
\label{eq:scalar_anchor}
\end{align}
For real reference CFFs, unpolarized data constrain the real scalar interference and $|{\cal S}|^2$, but do not determine the sign of the imaginary interference.  Since the present analysis addresses the separation of the scalar and GFF directions, we restrict $S_*$ to be real here and treat the physical relative phase separately in Sec.~\ref{sec:phase}.  The reference shape is $(M_S,p_S,\nu_S)=(0.705~\GeV,2,0)$.

One overall normalization factor is fitted for each experiment with its published normalization uncertainty included in the fit; the exact definition and parameter ranges are given in Appendix~\ref{app:fitrecords}.  For the reference ranges, the fit with $S_*=0$ gives $\chi^2=25.0150$.  Allowing the scalar term gives a positive solution
\[
 S_*=1.9221,\qquad \chi^2=24.5521,
\]
and a negative solution
\[
 S_*=-0.8450,\qquad \chi^2=24.6536.
\]
The two signs therefore describe the present unpolarized data almost equally well.  Several GFF parameters reach their allowed limits already for $S_*=0$, and enlarging those ranges moves the preferred scalar coefficient rather than stabilizing it.

\begin{figure}[tbp]
\centering
\includegraphics[width=\columnwidth]{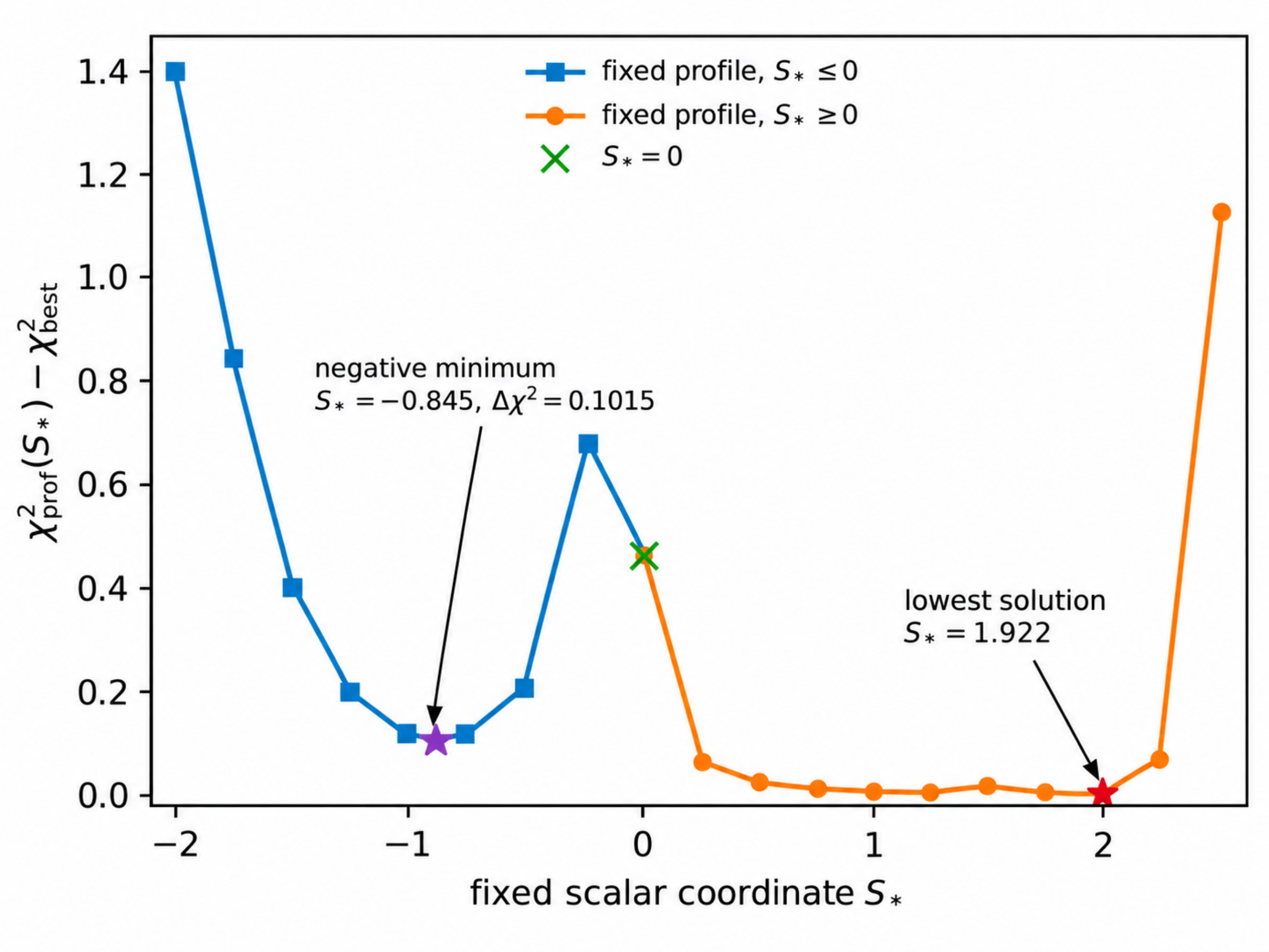}
\caption{$\chi^2$ as a function of $S_*$ in the reference real leading-CFF parametrization, with the six GFF parameters and experiment normalization factors fitted at each value of $S_*$.  The stars mark the positive and negative minima; their $\chi^2$ values differ by only $0.1015$.}
\label{fig:sstar_profile}
\end{figure}

The sensitivity to the assumed scalar profile and GFF range is illustrated by the representative cases in Table~\ref{tab:robustness_main}.  The fitted scalar coefficient changes substantially across these choices, while finite-bin averaging over the physical part of each bin with $\xi>0.5$ leaves the reference conclusion essentially unchanged.  The complete set of variations, parameter values and bin-averaging prescription are given in Appendix~\ref{app:fitrecords}.

\begin{table}[tbp]
\centering
\scriptsize
\caption{Representative variations of the differential fit.  $\chi^2_0$ denotes the fit with $S_*=0$, and $\chi^2_S$ the fit with $S_*$ varied.}
\label{tab:robustness_main}
\begin{ruledtabular}
\begin{tabular}{lrrr}
Variation & $\chi^2_0$ & $\chi^2_S$ & $S_*$\\
\hline
reference & 25.015 & 24.552 & 1.922\\
GYZ parameter range & 25.994 & 25.909 & 0.982\\
enlarged GFF range & 24.418 & 21.992 & -5.000\\
$\nu_S=+1$ & 25.015 & 19.683 & 2.304\\
finite-bin average, $\xi>0.5$ & 27.846 & 27.147 & 2.312\\
\end{tabular}
\end{ruledtabular}
\end{table}

Present unpolarized differential data therefore do not isolate a scalar production coefficient from the retained GFF direction in a profile-independent way.  Proton-helicity information can distinguish some of the resulting global fit solutions, although it does not by itself remove the exact local collinearity between ${\cal S}$ and ${\cal C}_{\rm eff}$.

\subsection{Helicity structure of the retained CFF directions}

The same fitted CFF solutions contain helicity information that is lost in the proton-spin sum of Eq.~\eqref{eq:gpd_bilinear}.  In symmetric GPD kinematics define
\begin{equation}
 t_0=-\frac{4M^2\xi^2}{1-\xi^2}.
\label{eq:t0_helicity}
\end{equation}
Up to the conventional azimuthal phase of the proton-helicity-flip amplitude, the two independent nucleon-helicity structures are~\cite{Diehl2003}
\begin{align}
 G_{++}&=\sqrt{1-\xi^2}\left({\cal H}-\frac{\xi^2}{1-\xi^2}{\cal E}\right),\notag\\
 G_{-+}&=-\frac{\sqrt{t_0-t}}{2M}{\cal E}.
\label{eq:gpd_helicity_amplitudes}
\end{align}
They satisfy the exact identity
\begin{equation}
 |G_{++}|^2+|G_{-+}|^2=|G|^2,
\label{eq:helicity_sum_closure}
\end{equation}
with the right-hand side given by Eq.~\eqref{eq:gpd_bilinear}.  Thus the unpolarized fit fixes only the sum of the two helicity intensities.

\begin{figure}[tbp]
\centering
\includegraphics[width=\columnwidth]{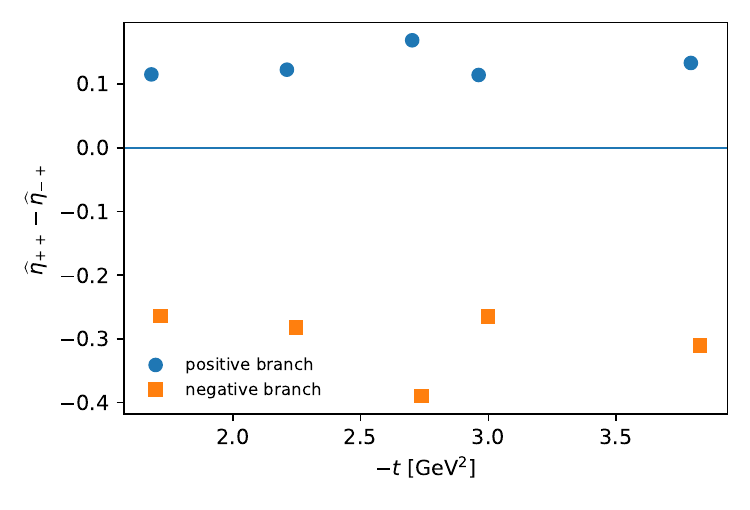}
\caption{CFF-level helicity contrast $\widehat\eta_{++}-\widehat\eta_{-+}$ at the five selected CLAS12 point centers.  The positive and negative solutions occupy opposite sides of zero, displaying the different proton-helicity content of the phenomenological scalar and retained GFF directions before process-specific photon--$J/\psi$ matching.}
\label{fig:helicity_contrast}
\end{figure}

Within the retained real leading-CFF representation, separate the phenomenological scalar direction from the retained GFF direction by
\begin{align}
 ({\cal H}_{S},{\cal E}_{S})&=({\cal S},-{\cal S}),\notag\\
 ({\cal H}_{G},{\cal E}_{G})&=({\cal H}_A+{\cal C}_{\rm eff},-{\cal C}_{\rm eff}).
\label{eq:cff_source_split}
\end{align}
The labels $S$ and $G$ denote the phenomenological scalar and retained GFF directions in this CFF representation; the operator sectors remain the $a=0,2$ decomposition defined in Sec.~\ref{sec:production}.  Equation~\eqref{eq:gpd_helicity_amplitudes} then gives
\begin{align}
 G_{S,++}&=\frac{{\cal S}}{\sqrt{1-\xi^2}}, &
 G_{G,++}&=\sqrt{1-\xi^2}{\cal H}_A+\frac{{\cal C}_{\rm eff}}{\sqrt{1-\xi^2}},\notag\\
 G_{S,-+}&=\frac{\sqrt{t_0-t}}{2M}{\cal S}, &
 G_{G,-+}&=\frac{\sqrt{t_0-t}}{2M}{\cal C}_{\rm eff},
\label{eq:source_helicity_amplitudes}
\end{align}
where the common flip-channel phase has been suppressed because it cancels in the ratios below.  For ${\cal S}\neq0$, and to distinguish these CFF-level ratios from the physical production ratios of Sec.~\ref{sec:production}, define
\begin{align}
 \widehat\eta_{++}&\equiv\frac{G_{G,++}}{G_{S,++}}
 =\frac{(1-\xi^2){\cal H}_A+{\cal C}_{\rm eff}}{{\cal S}},\notag\\
 \widehat\eta_{-+}&\equiv\frac{G_{G,-+}}{G_{S,-+}}
 =\frac{{\cal C}_{\rm eff}}{{\cal S}},
\label{eq:cff_helicity_ratios}
\end{align}
so that
\begin{equation}
 \boxed{\widehat\eta_{++}-\widehat\eta_{-+}
 =\frac{(1-\xi^2){\cal H}_A}{{\cal S}}}.
\label{eq:helicity_ratio_closure}
\end{equation}
Because both ratios scale as $1/{\cal S}$, their magnitude depends strongly on which fitted scalar solution is used and becomes singular as ${\cal S}\to0$.  Within the retained CFF decomposition, Eq.~\eqref{eq:helicity_ratio_closure} algebraically isolates the $A$-GFF contribution that is absent from the flip ratio.  At the reference point $(t_*,\xi_*)$, the positive solution gives $(\widehat\eta_{++},\widehat\eta_{-+})=(-0.834,-0.976)$, whereas the negative solution gives $(-1.378,-1.048)$.  Over the five CLAS12 points with $\xi>0.5$, the contrast in Eq.~\eqref{eq:helicity_ratio_closure} lies between $+0.114$ and $+0.169$ for the positive solution, but between $-0.390$ and $-0.264$ for the negative solution.  Thus the two solutions, whose unpolarized $\chi^2$ values are nearly equal, predict different proton-helicity compositions.  The opposite signs of the contrast can therefore discriminate between these fitted solutions.  They do not, however, assign the common $(1,-1)$ CFF direction separately to ${\cal S}$ and ${\cal C}_{\rm eff}$ at fixed kinematics.  Connecting this helicity difference to measured polarization observables further requires the process-dependent photon--$J/\psi$ helicity matching.  Figure~\ref{fig:helicity_contrast} shows the separation between the two solutions.

The contrast between the two fitted solutions therefore provides a possible polarization discriminator without lifting the fixed-kinematics ${\cal S}$--${\cal C}_{\rm eff}$ collinearity.  Mapping the CFF-level structures $G_{S,\lambda'\lambda}$ and $G_{G,\lambda'\lambda}$ onto SDMEs or spin asymmetries requires the process-dependent photon--$J/\psi$ helicity amplitudes of Sec.~\ref{sec:production}.

\section{Phase information and polarization observables}
\label{sec:phase}

The remaining phase question is dynamical.  For fixed $J^P$, collect the strongly coupled hadronic and $LS$/helicity components into a channel vector.  On the physical right-hand cut production unitarity can be written
\begin{equation}
 {\cal F}^{J,+}_{a,i}-{\cal F}^{J,-}_{a,i}
 =2i\sum_j T_{ij}^{J,-}\,\rho_j\,{\cal F}^{J,+}_{a,j},
 \qquad a=0,2,
\label{eq:matrix_unitarity_general}
\end{equation}
where $\rho$ is the diagonal phase-space matrix~\cite{Watson1952,Oller2020}.  If the projected production kernels are real on the elastic right-hand cut, so that they carry no independent elastic discontinuity, Watson's theorem gives the same right-hand-cut factor to the two components in one elastic eigenchannel,
\begin{equation}
 {\cal F}_{a,r}^{J}(s)=\Omega_{Jr}(s){\cal M}_{a,r}^{J}(s),
 \qquad \Omega_{Jr}=|\Omega_{Jr}|e^{i\delta_{Jr}},
\label{eq:watson_eigenchannel}
\end{equation}
so, away from production zeros,
\begin{equation}
 \eta_{{\rm phys},r}^{J}=\eta_{{\rm prod},r}^{J},
 \qquad
 \phi_{20,{\rm phys},r}^{J}=\phi_{20,{\rm prod},r}^{J}.
\label{eq:phase_preservation}
\end{equation}
Under this condition a common elastic final-state interaction cannot generate a new scalar--spin-two phase.

With several coupled channels the propagation is instead described schematically by a Muskhelishvili--Omn\`es matrix~\cite{Oller2020,Omnes1958},
\begin{equation}
 \bm{{\cal F}}_a^J(s)=\bm\Omega_J(s)\,\mathbf P_a^J(s).
\label{eq:matrixOmnes}
\end{equation}
If $\mathbf P_2^J=c(s)\mathbf P_0^J$, the common rescattering matrix preserves proportionality.  For real production vectors on the right-hand cut, a new strong rescattering phase therefore requires nonparallel production vectors, overlap with more than one strong eigenchannel, and different eigenphases in those channels.  Near-physical lattice QCD finds attractive $NJ/\psi$ interactions with moderate spin dependence in the $S$-wave scattering lengths~\cite{LyuDoiHatsudaSugiura2025}.  The lower $N\eta_c$ channel enters the $J^P=1/2^-$ sector, but $NJ/\psi\leftrightarrow N\eta_c$ changes the heavy-quark spin and starts with HQSS-breaking interactions such as chromomagnetic $1/m_c$ terms~\cite{Syamtomov2026Matching,BrambillaPinedaSotoVairo2000,SugiuraIkedaIshii2019}.  Open-charm channels add further rescattering paths~\cite{WinneyJPAC2023,Zhang2025,ClymtonKimKim2026,SakinahKimChoi2026}.

To determine the sign of a relative phase, one needs an imaginary interference quadrature in addition to the real bilinears fixed by an unpolarized rate.  The mapping of a measured polarization response onto that quadrature is fixed by the production helicity amplitudes.  Near-threshold quarkonium electroproduction has been studied as a probe of gluonic structure~\cite{BoussarieHatta2020}, and polarization observables have been proposed to distinguish production mechanisms~\cite{WinneyEtAl2019}.  Up to the virtual-photon flux and convention-dependent signs, the azimuthal dependence for an unpolarized target has the standard form~\cite{SchillingWolf1973}
\begin{align}
 d\sigma \propto{}& \sigma_T+\epsilon\sigma_L
 +\epsilon\sigma_{TT}\cos2\varphi
 +\sqrt{2\epsilon(1+\epsilon)}\,\sigma_{LT}\cos\varphi\notag\\
 &+h\sqrt{2\epsilon(1-\epsilon)}\,\sigma_{LT'}\sin\varphi,
\label{eq:electroproduction_responses}
\end{align}
where the helicity-odd $LT'$ response contains imaginary longitudinal--transverse interference.  In the full reaction it is a helicity-weighted sum,
\begin{equation}
 \sigma_{LT'}\;\propto\;\sum_{\alpha}w_{\alpha}\,
 \mathrm{Im}\!\left[\mathcal A_{L,\alpha}^*\mathcal A_{T,\alpha}\right],
\label{eq:LT_helicity_sum}
\end{equation}
where $\alpha$ collects the nucleon and vector-meson helicities; the weights and relative signs depend on convention.  The $J/\psi\to\ell^+\ell^-$ decay distribution supplies further spin-density-matrix bilinears, as already demonstrated in exclusive $J/\psi$ electroproduction at HERA~\cite{ZEUS2004}.  The Bayesian amplitude framework of ~\cite{SinghEtAl2026Amplitude}, developed for lighter vector mesons, shows how polarization and decay-angle information can be propagated to helicity amplitudes while retaining discrete ambiguities.

The production matching must be retained explicitly before an imaginary interference can be assigned to the scalar--spin-two phase.  For one reduced factorized two-component amplitude, with the helicity label $\alpha$ suppressed, let
\[
 {\cal F}_{a,L}=c_{aL}X_a,\qquad
 {\cal F}_{a,T}=c_{aT}X_a,\qquad a=0,2,
\]
then $\mathcal A_L=c_{0L}X_0+c_{2L}X_2$ and $\mathcal A_T=c_{0T}X_0+c_{2T}X_2$.  This factorization assumes that the same two source coordinates $X_0$ and $X_2$ generate both reduced longitudinal and transverse amplitudes.  If the projected matching coefficients are real,
\begin{equation}
 \mathrm{Im}({\cal A}_L^*{\cal A}_T)
 =\big(c_{0L}c_{2T}-c_{2L}c_{0T}\big)
 \mathrm{Im}(X_0^*X_2).
\label{eq:LT_determinant}
\end{equation}
For general complex matching coefficients, writing $Z=X_0^*X_2$ gives
\begin{align}
 \mathrm{Im}({\cal A}_L^*{\cal A}_T)={}&
 \mathrm{Im}(c_{0L}^*c_{0T})|X_0|^2
 +\mathrm{Im}(c_{2L}^*c_{2T})|X_2|^2 \notag\\
 &+\mathrm{Re}\!\left(c_{0L}^*c_{2T}-c_{2L}^*c_{0T}\right)\mathrm{Im}Z \notag\\
 &+\mathrm{Im}\!\left(c_{0L}^*c_{2T}+c_{2L}^*c_{0T}\right)\mathrm{Re}Z .
\label{eq:LT_general_complex}
\end{align}
Equation~\eqref{eq:LT_general_complex} separates the sources of the imaginary longitudinal--transverse bilinear.  Diagonal matching phases contribute terms proportional to $|X_a|^2$, while complex off-diagonal matching also brings $\mathrm{Re}Z$ into the response.  Only in the real two-coefficient reduction do these terms vanish; Eq.~\eqref{eq:LT_determinant} then gives the condition for sensitivity to $\mathrm{Im}(X_0^*X_2)$,
\[
 c_{0L}c_{2T}-c_{2L}c_{0T}\neq0.
\]
If the same two source coordinates factorize across the helicity sum in Eq.~\eqref{eq:LT_helicity_sum}, define
\begin{align}
 D_a&=\sum_\alpha w_\alpha\,\mathrm{Im}(c_{aL,\alpha}^*c_{aT,\alpha}),\qquad a=0,2,\notag\\
 K_I&=\sum_\alpha w_\alpha\,\mathrm{Re}\left(c_{0L,\alpha}^*c_{2T,\alpha}-c_{2L,\alpha}^*c_{0T,\alpha}\right),\notag\\
 K_R&=\sum_\alpha w_\alpha\,\mathrm{Im}\left(c_{0L,\alpha}^*c_{2T,\alpha}+c_{2L,\alpha}^*c_{0T,\alpha}\right).
\label{eq:LT_helicity_coefficients}
\end{align}
The physical response then has the form
\begin{equation}
 \sigma_{LT'}\propto D_0|X_0|^2+D_2|X_2|^2+K_I\,\mathrm{Im}Z+K_R\,\mathrm{Re}Z.
\label{eq:LT_physical_factorized}
\end{equation}
For real matching coefficients $D_0=D_2=K_R=0$, and phase sensitivity requires the helicity-weighted determinant sum $K_I\neq0$.  For complex coefficients the diagonal and $\mathrm{Re}Z$ terms must be determined together with the desired imaginary interference.  If the two-source factorization fails across helicity channels, Eq.~\eqref{eq:LT_helicity_sum} must instead be evaluated with the full process-specific amplitudes.  A scalar--spin-two phase extraction then requires enough independent polarization responses for the corresponding bilinear coefficient map to have sufficient rank.  Target, recoil and vector-meson polarization observables provide additional such bilinears.

\begin{figure*}[t]
\centering
\includegraphics[width=0.48\textwidth]{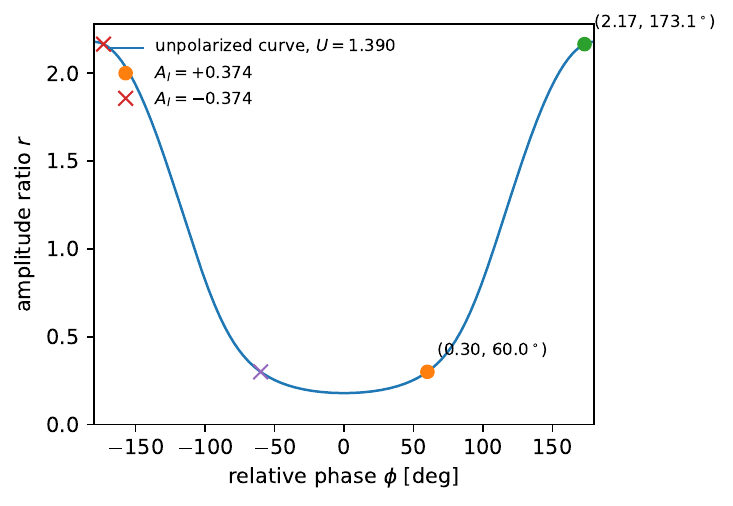}
\caption{Two-complex-amplitude geometry for ${\cal F}_2=1$.  The unpolarized coordinate $U=1.390$ defines the curve; filled points satisfy $A_I=+0.374$, while crosses show the mirror solutions with $A_I=-0.374$.  In electroproduction the coefficient of the corresponding imaginary interference is determined by the helicity matching factors in Eqs.~\eqref{eq:LT_helicity_coefficients} and~\eqref{eq:LT_physical_factorized}.}
\label{fig:phase_geometry}
\end{figure*}

A two-amplitude example illustrates the information carried by an independent imaginary quadrature once this mapping is specified.  Choose a known reference amplitude ${\cal F}_2=1$ and write ${\cal F}_0=r e^{i\phi}$.  The unpolarized coordinate
\[
 U=|{\cal F}_2+{\cal F}_0|^2=1+r^2+2r\cos\phi
\]
defines a continuous curve and is invariant under $\phi\to-\phi$.  A second coordinate proportional to the missing imaginary interference,
\[
 A_I=\frac{2\,\mathrm{Im}({\cal F}_2^*{\cal F}_0)}{U}
 =\frac{2r\sin\phi}{U},
\]
selects isolated points on that curve.  For the illustrative choice $U=1.390$ and $A_I=0.374$, one solution is $(r,\phi)=(0.30,60^\circ)$ and the second is approximately $(2.17,173.1^\circ)$.  Figure~\ref{fig:phase_geometry} shows the resulting algebraic geometry: the unpolarized coordinate leaves a continuous ambiguity, while an independent imaginary interference selects two discrete branches.

Complementary kinematic separations have also been proposed for large-$|t|$ threshold production~\cite{PentchevChudakov2025}.  The published 2026 CLAS12 result used in the present 38-point fit is an unpolarized total- and differential-cross-section measurement~\cite{CLAS122026}; it does not provide the polarization or separated longitudinal/transverse information required by Eqs.~\eqref{eq:LT_helicity_sum}--\eqref{eq:LT_physical_factorized}.  Separate CLAS12 analyses reported at MESON2026 are still in progress.  A detected-electron free-proton analysis is in progress and is designed to probe roughly $0.01<Q^2<0.15~\GeV^2$, with the stated aim of placing initial constraints on the $Q^2$ dependence, while the polarization program aims to measure unpolarized spin-density matrix elements, beam-spin asymmetries, longitudinal target-spin asymmetries and beam--target double-spin asymmetries~\cite{TysonMESON2026}.  Longitudinally polarized $\mathrm{NH}_3$ data were taken in 2022--23, but the status report describes the polarization analysis as in progress/early stage.  These ongoing analyses are therefore not inputs to the present fit; once available, they can provide additional helicity bilinears with which to test the solution-dependent pattern in Eq.~\eqref{eq:helicity_ratio_closure}, subject to the scalar and spin-two photon--$J/\psi$ matching.

The approved $\mu$CLAS12 program is a future high-luminosity extension intended to provide high-statistics $J/\psi$ electroproduction at $Q^2>0$~\cite{AlvaradoMuCLAS2026,TysonMESON2026}.  Jefferson Lab experiment E12-12-006 is likewise an approved future setting for near-threshold electroproduction~\cite{SoLIDJpsi}.  Interpreting their measured $Q^2$ dependence and polarization asymmetries in terms of $\widehat\eta_{++}$, $\widehat\eta_{-+}$ or a scalar--spin-two phase will require the corresponding process-specific matching.  Information on $N\eta_c$ and open-charm transitions would further constrain the final-state sector and coupled-channel phases.

\section{Conclusions}

The integrated and differential analyses lead to the same conclusion
by different routes: present unpolarized near-threshold data do not
provide a unique separation of a scalar production contribution from
the other amplitude structures with which it is correlated.  This
statement concerns the production amplitude.  Relating a fitted
phenomenological scalar contribution to the source-level scalar
response $N_0$, or a corresponding spin-two contribution to $N_2$,
requires the process-dependent matching discussed in
Sec.~\ref{sec:production}.

For the nine GlueX-2023 points with $W\le4.55~\GeV$, adding a scalar
term only to the Born production amplitude changes
$\chi^2_{\rm int}$ from $49.185$ to $45.428$.  With the elastic
$K$-matrix denominator included, the corresponding change is
$51.561\to46.278$.  A common scalar coefficient in production and
elastic rescattering gives a substantially larger improvement for the
fixed steep non-forward profile, reaching
$\chi^2_{\rm int}=28.228$.  This preference is not stable, however:
when the production and elastic scalar coefficients are varied
independently they become strongly correlated, while harder
non-forward elastic profiles describe the energy dependence without
a nonzero shared scalar contribution.  The integrated data therefore
determine neither a unique scalar production strength, its allocation
to elastic rescattering, nor a scalar length scale.

Differential measurements retain the $t$ dependence but do not remove
the structural degeneracy.  In the 38-point real leading-CFF
large-$\xi$ calculation, the phenomenological scalar term and the
retained $C$-GFF combination occupy exactly the same
$(\delta{\cal H},\delta{\cal E})\propto(1,-1)$ direction at fixed
kinematics.  They can therefore be distinguished across the measured
sample only through their assumed $(t,\xi)$ dependences.  For the
reference scalar profile the positive solution lowers $\chi^2$ by
only $0.4629$ relative to $S_*=0$, while the negative solution lies
only $0.1015$ higher.  Both the fitted value of $S_*$ and the apparent
improvement change substantially when the scalar profile or the
allowed GFF ranges are varied, and the preferred scalar value does not
stabilize as those ranges are enlarged.  Finite-bin averaging leaves
this qualitative conclusion unchanged.  Present unpolarized
differential data therefore do not isolate a scalar production
coefficient from the retained GFF direction in a profile-independent
way.

The proton-helicity decomposition nevertheless distinguishes the
helicity content of the two nearly degenerate global fit solutions.
At the five selected CLAS12 points, the CFF-level contrast
$\widehat\eta_{++}-\widehat\eta_{-+}$ has opposite sign for the
positive and negative scalar solutions.  This provides a possible
polarization discriminator between the fitted solutions, but it does
not remove the exact local ${\cal S}$--${\cal C}_{\rm eff}$
collinearity.  Connecting this contrast to measurable SDMEs or spin
asymmetries requires the process-dependent photon--$J/\psi$ helicity
amplitudes.

The relative phase introduces an additional dynamical requirement.
For real projected production kernels in a single elastic
eigenchannel, Watson's theorem gives the scalar and spin-two
components the same final-state factor.  Their ratio and relative
phase are therefore preserved.  Coupled-channel rescattering can
generate a new relative phase only when the two production vectors
are nonparallel and populate at least two strong eigenchannels with
different eigenphases.

Finally, unpolarized cross sections contain modulus-squared and
real-interference combinations but do not determine the signed
imaginary interference needed to fix the relative phase.  In the
reduced real-coefficient two-source limit, a longitudinal--transverse
response is sensitive to this quadrature only when the corresponding
matching determinant is nonzero; with general complex matching,
additional diagonal and real-interference terms must be determined as
well.  The presently published CLAS12 cross sections do not provide
these independent constraints.  Polarization-sensitive measurements,
together with process-specific complex helicity amplitudes, are
therefore required for a scalar--spin-two phase determination.

\section*{Data availability}
The experimental inputs are given in~\cite{GlueX2023,Duran2023,CLAS122026,GlueXHEPData2023}.  The numerical fit uses the published tables available as of 6 September 2026.  The later $J/\psi$-007 dimuon/combined two-dimensional result~\cite{Jpsi007Muon2026} is discussed for comparison but is not included because a corresponding point-by-point two-dimensional cross-section table is not publicly available.  The numerical tables and programs used for the integrated fits, together with the data underlying the figures, are provided as supplementary material.  The supplement also contains full-precision fit results, point-by-point residuals and the detailed numerical checks used to verify the calculations.  The finite-bin calculation uses a uniform average over the intersection of each published bin with exact two-body support and $\xi>0.5$; an acceptance-weighted average would require common event-level acceptance information for all three differential data sets.

\appendix
\section{Integrated-fit details}
\label{app:integrated_representation}

The integrated analysis can be expressed directly in terms of three calculated cross sections: with the scalar term omitted, with a scalar term added at Born level, and with a common scalar coefficient in the production and elastic terms.  Let these be $\sigma_i^{\rm II}$, $\sigma_i^{\rm III}$ and $\sigma_i^{\rm IV}$, with the common scalar coefficient $C_0=6.45924$.  With $C_P\equiv C_S^{(P)}$ and $C_K\equiv C_S^{(K)}$,
\begin{equation}
 B_i=\sqrt{\sigma_i^{\rm II}},\qquad
 Q_i=\frac{\sqrt{\sigma_i^{\rm III}}-B_i}{C_0},
\label{eq:response_reconstruction_BQ}
\end{equation}
so that $\sigma_i^{\rm Born}(C_P)=(B_i+C_PQ_i)^2$.  The shared-$K$ result determines
\begin{equation}
 |D_i|=\frac{1}{\rho_i}
 \sqrt{\frac{\sigma_i^{\rm III}}{\sigma_i^{\rm IV}}-1},
 \qquad D_i=K_i+C_0\bar G_{S,i},
\label{eq:response_reconstruction_D}
\end{equation}
and hence
\begin{equation}
 K_i^{(\pm)}=\pm|D_i|-C_0\bar G_{S,i}.
\label{eq:response_reconstruction_Kpm}
\end{equation}
For arbitrary production and elastic scalar coefficients,
\begin{equation}
 \sigma_i(C_P,C_K;\pm)=
 \frac{(B_i+C_PQ_i)^2}
 {1+\rho_i^2[K_i^{(\pm)}+C_K\bar G_{S,i}]^2}.
\label{eq:response_reconstruction_sigma}
\end{equation}
The two coherent sign choices give identical cross sections under
\begin{equation}
 C_K^{(-)}=2C_0-C_K^{(+)}.
\label{eq:D_sign_map}
\end{equation}
Thus $C_K=0$ in the positive-$D$ convention maps to $C_K=2C_0$ in the opposite convention.

At fixed normalization the independent fit gives
\[
 (C_P,C_K)=(61.1423,22.2630),\qquad \chi^2_{\rm int}=0.50009.
\]
When an overall normalization is also fitted the minimum moves to
\begin{align*}
 C_P&=124.8320,\qquad C_K=8.4749,\\
 n_{\rm int}&=0.40643,\qquad \chi^2=0.38306.
\end{align*}
The fitted coefficients remain at the edge of the allowed range until that range is extended substantially: with fixed normalization the $C_P$ limit persists through $|C_{P,K}|=50$, while with $n_{\rm int}$ fitted it persists through $|C_{P,K}|=100$.  The local correlation changes from $0.917$ at fixed normalization to $-0.979$ when $n_{\rm int}$ is fitted, and the smaller Hessian eigenvalue decreases by about two orders of magnitude.

To compare the energy dependence for different non-forward kernels we fit one overall normalization for each profile,
\begin{equation}
 \chi^2_{\rm shape}(\theta)=\min_{n_{\rm int}}\sum_{i=1}^{9}
 \frac{[\sigma_i^{\rm data}-n_{\rm int}\sigma_i^{\rm model}(\theta)]^2}{(\delta\sigma_i)^2}.
\label{eq:integrated_shape_stat}
\end{equation}
For the steep profile, fitting only $n_{\rm int}$ gives $\chi^2_{\rm shape}=49.162$, $43.049$ and $39.098$ for the curves with no scalar term, a Born scalar term and a common production/elastic scalar coefficient, respectively.  A constant-slope exponential gives $b_0=0.240~\GeV^{-2}$ and $\chi^2_{\rm shape}=0.408$.  Fixed dipoles give $3.062$ for $\Lambda=2~\GeV$ and $0.531$ for $\Lambda=M_\psi$; fitting the dipole scale gives $\Lambda=3.328~\GeV$ and $\chi^2_{\rm shape}=0.457$.  A linear slope $b(W)=b_0+b_1(W-W_{\rm th})$ gives $b_0=0.671~\GeV^{-2}$, $b_1=0.00429~\GeV^{-3}$ and $\chi^2=0.562$.  In each of these harder-profile fits the shared scalar coefficient is driven to zero.

In the corresponding scalar-mass scan the fit becomes nearly flat for $m_s\gtrsim3~\GeV$, with $\Delta\chi^2_{\rm shape}=1$ near $m_s=3.26~\GeV$.  The cross-section representation reconstructed for the steep profile is profile specific: rescaling it does not reproduce the directly calculated dipole and constant-slope results, so independent $(C_P,C_K)$ values for harder continuations require their directly calculated cross sections.

\section{Differential-fit details}
\label{app:fitrecords}

One normalization factor $n_e$ is fitted for each experiment, with widths $19.5\%$ for GlueX, $4.0\%$ for $J/\psi$-007 and $11.198\%$ for CLAS12.  For CLAS12 the quoted scale component is removed in quadrature from the published total systematic error before the same scale is included through $n_e$; for GlueX and $J/\psi$-007 the published scale errors are additional to the pointwise systematic columns.  The fit function is
\begin{equation}
 \chi^2_{\rm prof}=\sum_e\sum_{i\in e}
 \frac{(d_i-n_e m_i)^2}{\sigma_i^2}
 +\sum_e\frac{(n_e-1)^2}{\delta_{e,{\rm norm}}^2},
\label{eq:differential_chi2}
\end{equation}
with
\begin{equation}
 \widehat n_e=
 \frac{\displaystyle\sum_{i\in e}d_i m_i/\sigma_i^2
       +\delta_{e,{\rm norm}}^{-2}}
      {\displaystyle\sum_{i\in e}m_i^2/\sigma_i^2
       +\delta_{e,{\rm norm}}^{-2}}
\label{eq:analytic_norm_profile}
\end{equation}
for fixed model parameters.  The covariance contains the published point uncertainties and these three correlated normalization terms; the GYZ theory covariance~\cite{GuoYuanZhao2025} is not included.

For the reference scalar shape $(M_S,p_S,\nu_S)=(0.705~\GeV,2,0)$ we use $0.05\le M_{A_i},M_{C_i}\le8~\GeV$, $|C_{q,g}(0)|\le8$ and $|S_*|\le5$.  The GYZ parameter range is $0\le M_{A_i},M_{C_i}\le4~\GeV$ and $|C_{q,g}(0)|\le4$, with the zero-mass endpoint implemented numerically as $M_i\ge10^{-3}~\GeV$.  Enlarging the ranges to $0.01\le M_i\le20~\GeV$, $|C_i(0)|\le20$ and $|S_*|\le5$ gives $\chi^2_0=24.41819$ and $\chi^2_S=21.99243$ at $S_*=-5$ and $M_{A_q}=0.01~\GeV$.  Extending the scalar range to $|S_*|\le10$ lowers $\chi^2_S$ to $21.83401$ at $S_*\simeq-5.56$, and increasing the allowed $C_g$ range lowers it further.

\begin{table*}[tbp]
\centering
\footnotesize
\setlength{\tabcolsep}{5.2pt}
\begin{tabular}{lccc}
\hline
Quantity & $S_*=0$ & positive solution & negative solution \\
\hline
$S_*$ & $0$ & $1.9221$ & $-0.8450$ \\
$\chi^2_{\rm prof}$ & $25.0150$ & $24.5521$ & $24.6536$ \\
$\chi^2_{\rm prof}-\chi^2_{\rm best}$ & $0.4629$ & $0$ & $0.1015$ \\
$M_{A_g}$ [GeV] & $3.258$ & $3.224$ & $3.244$ \\
$C_g(0)$ & $-2.477$ & $-7.094$ & $8.000$ \\
$M_{C_g}$ [GeV] & $2.721$ & $1.208$ & $0.862$ \\
$M_{A_q}$ [GeV] & $8.000$ & $8.000$ & $8.000$ \\
$C_q(0)$ & $-8.000$ & $0.871$ & $-0.506$ \\
$M_{C_q}$ [GeV] & $2.814$ & $2.154$ & $3.452$ \\
$n_{\rm GlueX}$ & $1.0658$ & $1.0857$ & $1.0743$ \\
$n_{J/\psi\text{-}007}$ & $0.9957$ & $0.9962$ & $0.9960$ \\
$n_{\rm CLAS12}$ & $1.0342$ & $1.0241$ & $1.0296$ \\
$\chi^2_{\rm GlueX}$ & $10.127$ & $8.973$ & $9.521$ \\
$\chi^2_{J/\psi\text{-}007}$ & $12.004$ & $12.283$ & $12.122$ \\
$\chi^2_{\rm CLAS12}$ & $2.884$ & $3.296$ & $3.011$ \\
parameters at limits & $M_{A_q}\!\uparrow,C_q(0)\!\downarrow$
 & $M_{A_q}\!\uparrow$ & $C_g(0)\!\uparrow,M_{A_q}\!\uparrow$ \\
\hline
\end{tabular}
\caption{Parameters and experiment contributions for the reference differential fit.  Arrows indicate parameters at their upper or lower allowed values.}
\label{tab:reference_solutions}
\end{table*}

The complete set of variations used to test the scalar profile, GFF ranges and data selection is listed in Table~\ref{tab:robustness}.  The normalization factors are minimized analytically at each point.

\begin{table*}[tbp]
\caption{Dependence of the differential fit on the allowed GFF range, scalar profile and data selection.  $\chi^2_0$ denotes $S_*=0$ and $\chi^2_S$ the fit with $S_*$ varied.  Arrows indicate parameters at their allowed limits.}
\label{tab:robustness}
\begin{ruledtabular}
\begin{tabular}{lrrrrll}
Variation & $N$ & $\chi^2_0$ & $\chi^2_S$ & $\Delta\chi^2$ & $S_*$ & parameters at limits ($S_*=0$; $S_*$ fitted) \\
\hline
reference & 38 & 25.015 & 24.552 & 0.463 & 1.922 & $M_{A_q}\uparrow,C_q(0)\downarrow$; $M_{A_q}\uparrow$ \\
GYZ parameter range & 38 & 25.994 & 25.909 & 0.085 & 0.982 & --; $C_g(0)\downarrow,M_{A_q}\downarrow$ \\
enlarged GFF range & 38 & 24.418 & 21.992 & 2.426 & -5.000 & $M_{A_q}\uparrow$; $M_{A_q}\downarrow,S_*\downarrow$ \\
$p_S=1$ & 38 & 25.015 & 23.550 & 1.465 & 2.654 & $M_{A_q}\uparrow,C_q(0)\downarrow$; $C_g(0)\downarrow,M_{A_q}\downarrow$ \\
$p_S=3$ & 38 & 25.015 & 23.952 & 1.063 & -0.545 & $M_{A_q}\uparrow,C_q(0)\downarrow$; $C_g(0)\uparrow,M_{A_q}\downarrow,C_q(0)\uparrow$ \\
$M_S=1.20~\GeV$ & 38 & 25.015 & 23.903 & 1.113 & -5.000 & $M_{A_q}\uparrow,C_q(0)\downarrow$; $M_{A_q}\downarrow,M_{C_q}\uparrow,S_*\downarrow$ \\
$\nu_S=-1$ & 38 & 25.015 & 22.018 & 2.997 & 0.573 & $M_{A_q}\uparrow,C_q(0)\downarrow$; $M_{A_q}\uparrow$ \\
$\nu_S=-0.5$ & 38 & 25.015 & 22.507 & 2.508 & 1.053 & $M_{A_q}\uparrow,C_q(0)\downarrow$; $M_{A_q}\uparrow$ \\
$\nu_S=+0.5$ & 38 & 25.015 & 23.452 & 1.563 & -0.792 & $M_{A_q}\uparrow,C_q(0)\downarrow$; $M_{A_q}\uparrow$ \\
$\nu_S=+1$ & 38 & 25.015 & 19.683 & 5.332 & 2.304 & $M_{A_q}\uparrow,C_q(0)\downarrow$; $M_{A_q}\downarrow,M_{C_q}\uparrow$ \\
finite-bin average, $\xi>0.5$, $q=12$ & 38 & 27.846 & 27.147 & 0.698 & 2.312 & $M_{A_q}\downarrow,C_q(0)\uparrow$; $C_g(0)\downarrow,M_{A_q}\uparrow$ \\
$\xi>0.55$ & 24 & 10.133 & 8.603 & 1.530 & 4.222 & $C_g(0)\uparrow,M_{A_q}\downarrow,C_q(0)\uparrow$; $C_g(0)\downarrow,M_{A_q}\downarrow$ \\
without CLAS12 & 33 & 21.704 & 20.939 & 0.765 & 2.313 & $M_{A_q}\downarrow,C_q(0)\uparrow$; $C_g(0)\downarrow,M_{A_q}\uparrow$ \\
\end{tabular}
\end{ruledtabular}
\end{table*}

For the finite-bin calculation the theory is averaged over the intersection of each published bin with exact two-body support and $\xi>0.5$,
\begin{equation}
 \bar m_i=\frac{1}{\Omega_i}\int_{{\cal R}_i}dy_1\,dy_2\,m(y_1,y_2),
 \qquad
 \Omega_i=\int_{{\cal R}_i}dy_1\,dy_2,
\label{eq:finite_bin_average}
\end{equation}
where ${\cal R}_i=B_i\cap{\cal P}\cap\{\xi>0.5\}$.  Seven selected bins straddle $\xi=0.5$.  A $12\times12$ Gauss--Legendre rule gives $\chi^2_0=27.846$, $\chi^2_S=27.147$ and $S_*=2.312$.  Increasing the order to $q=20,24,28$ and $32$ gives $\Delta\chi^2=0.76,0.73,0.73$ and $0.73$, while $S_*$ remains between $2.307$ and $2.309$.

Finally, Table~\ref{tab:fitvectors} gives the parameter vectors for all variations.  We use the ordering
\begin{equation}
 \boldsymbol\theta=\paramvec,
\end{equation}
with masses in GeV.  The column with fitted $S_*$ appends it as the seventh entry.  Full-precision parameters and fitted normalizations are provided in the supplementary material.

\begin{table*}[tbp]
\scriptsize
\setlength{\tabcolsep}{3pt}
\caption{Parameter vectors for the differential fits.  The first column has $S_*=0$ fixed; the second lists $(\boldsymbol\theta;S_*)$ with the scalar coefficient fitted.}
\label{tab:fitvectors}
\begin{ruledtabular}
\begin{tabular}{lll}
Variation & $\boldsymbol\theta_0$ & $(\boldsymbol\theta_{S};S_*)$ \\
\hline
reference & $(3.258,-2.477,2.721,8.000,-8.000,2.814)$ & $(3.224,-7.094,1.208,8.000,0.871,2.154;1.922)$ \\
GYZ parameter range & $(1.620,-1.239,2.199,0.002,-3.470,2.405)$ & $(1.613,-4.000,1.231,0.001,0.091,2.313;0.982)$ \\
enlarged GFF range & $(3.498,-3.823,2.824,20.000,-12.572,2.889)$ & $(2.321,17.899,1.231,0.010,-0.275,6.142;-5.000)$ \\
$p_S=1$ & $(3.258,-2.477,2.721,8.000,-8.000,2.814)$ & $(2.326,-8.000,0.837,0.050,2.568,3.580;2.654)$ \\
$p_S=3$ & $(3.258,-2.477,2.721,8.000,-8.000,2.814)$ & $(2.344,8.000,1.276,0.050,8.000,1.778;-0.545)$ \\
$M_S=1.20~\GeV$ & $(3.258,-2.477,2.721,8.000,-8.000,2.814)$ & $(2.334,5.903,1.660,0.050,-0.262,8.000;-5.000)$ \\
$\nu_S=-1$ & $(3.258,-2.477,2.721,8.000,-8.000,2.814)$ & $(2.566,-2.994,1.223,8.000,-0.323,3.365;0.573)$ \\
$\nu_S=-0.5$ & $(3.258,-2.477,2.721,8.000,-8.000,2.814)$ & $(2.616,-4.886,1.199,8.000,-0.138,3.862;1.053)$ \\
$\nu_S=+0.5$ & $(3.258,-2.477,2.721,8.000,-8.000,2.814)$ & $(2.764,7.463,0.820,8.000,-0.753,3.194;-0.792)$ \\
$\nu_S=+1$ & $(3.258,-2.477,2.721,8.000,-8.000,2.814)$ & $(2.316,-6.579,1.404,0.050,0.155,8.000;2.304)$ \\
finite-bin, $q=12$ & $(2.381,2.714,1.667,0.050,8.000,1.880)$ & $(3.209,-8.000,1.263,8.000,0.244,3.142;2.312)$ \\
$\xi>0.55$ & $(2.159,8.000,1.060,0.050,8.000,1.604)$ & $(2.027,-8.000,1.548,0.050,0.219,5.965;4.222)$ \\
without CLAS12 & $(2.322,2.649,1.689,0.050,8.000,1.886)$ & $(3.170,-8.000,1.270,8.000,0.166,3.607;2.313)$ \\
\end{tabular}
\end{ruledtabular}
\end{table*}

\begingroup\sloppy

\endgroup

\end{document}